# MODELING INTER-COUNTRY CONNECTION FROM GEOTAGGED NEWS REPORTS: A TIME-SERIES ANALYSIS

*Yihong Yuan*

Department of Geography, Texas State University, San Marcos, TX, USA, 78666.

## ABSTRACT

The development of theories and techniques for big data analytics offers tremendous flexibility for investigating large-scale events and patterns that emerge over space and time. In this research, we utilize a unique open-access dataset "The Global Data on Events, Location and Tone" (GDELT) to model the image of China in mass media, specifically, how China has related to the rest of the world and how this connection has evolved upon time based on an autoregressive integrated moving average (ARIMA) model. The results of this research contribute in both methodological and empirical perspectives: We examined the effectiveness of time series models in predicting trends in long-term mass media data. In addition, we identified various types of connection strength patterns between China and its top 15 related countries. This study generates valuable input to interpret China's diplomatic and regional relations based on mass media data, as well as providing methodological references for investigating international relations in other countries and regions in the big data era.

## 1. INTRODUCTION

In recent decades, the rapid development of techniques and theories in the big data field has introduced new challenges and opportunities to analyze the large amount of information available online [1-3], including user-contributed (personalized) information such as social media data and central-generated traditional mass media. Social media are best characterized by a series of Social Network Sites (SNS) (such as Facebook and Twitter) that have attracted worldwide users to communicate, socialize and share their daily lives, whereas mass media refers to various forms of media technologies that aims to reach a large audience via mass communication, including broadcast, printed media, film, and new channels developed with the widespread of the world wide web (WWW), such as online news reports. Although many studies have focused on how the user-generated contents have revolutionized traditional media communications landscape especially in the marketing field [4-5], researchers have also realized the advantages of mass media contents in its professorial nature: Compared to social media, traditional mass media naturally addresses significant and aggregated events [6]; therefore, they play an important role in analyzing the social, economic, and cultural status of a society. In addition, traditional mass media has evolved for decades (or even centuries), and the data are often collected in a longer time span, so they are more appropriate to investigate long-term trends and patterns, such as the evolvement of an urban system in the past few decades, as well as exploring collective patterns of a society or the connection between societal systems [7].

Realizing the necessity to explore the geographic component of these geotagged news reports, this research utilizes an open-source dataset "The Global Data on Events, Location and Tone" (GDELT) to analyze the time series of inter-country connections of China with respect to time. GDELT monitors print, broadcast and web news media in over 100 languages worldwide and automatically encode such data into a structured database (CAMEO[1]-coded). Although researchers in various fields such as sociology and communication have explored the potential of such data in analyzing societal events [8-9], there is limited research in utilizing these extracted mass media data in geography, such as analyzing the evolvement of a geographic entity, or the connection between geographic entities upon time [10-11]. We adopt an autoregressive integrated moving average (ARIMA) model to analyze time series due to its capability of dealing with both Autoaggressive (AR) moving average (MA) and "Integrated" components. These models are appropriate for time series data either to better interpret the autocorrelation of the data, or to forecast future points in the series [12]. Additionally the ARIMA model has the capability to deal with non-stationary time series data which is typically associated with long-term news events. This research concentrates on demonstrating the effectiveness of applying time series analysis to geotagged mass media data. We do not aim to interpret these patterns from a sociological perspective. The applied methodology can be further extended to various fields as a data pre-processing strategy, such as public relation, communication and political geography.

---

[1] Conflict and Mediation Event Observations (CAMEO) is a framework for coding event data

## 2. METHODOLOGY

This research utilizes an open dataset named GDELT. This CAMEO-coded dataset[2] [13] is updated daily and consists of over a quarter-billion news event records dating back to 1979. It captures what has happened/is happening worldwide [3, 8]. The data include multiple columns such as the source, actors, time, and approximated location of recorded events. For instance, in news report entitled "In Malaysia, Obama carefully calibrates message to Beijing". Actor 1 will be "United States government" and Actor 2 is "Chinese Government". The associated geographic locations of Actor 1, Actor 2 and the actual action are "Beijing, China", "Washington DC, United States", and "Kuala Lumpur, Malaysia".

As discussed in Section 2, this research concentrates on the inter-country relatedness between China and foreign countries. The analyses will be conducted from the following two steps:

- *Data Preprocessing*

First we extract all news records involving China and another country as two parties. Note that the location of "action" is not considered as a substantial factor here, since an event related to a certain country can happen inside or outside of that country. Based on the pre-processed data, we calculate descriptive statistics to provide a general interpretation of the trend at various spatio-temporal scales. For each year each country, we calculate the frequencies of "co- occurrence" with China (donated as $C$) in the dataset. The frequencies are noted as $F_y(i,c)$, which stands for the "co-occurrence" frequency between China and country $i$ in year $y$. Here we first define connection strength as follows:

$$Co_y(i,c) = \frac{F_y(i,c)}{\sum_{j \neq c} F_y(j,c)} \quad (1)$$

Where $F_y(i,c)$ is the frequency of co-occurrence between China and $I$, and $\sum_{j \neq c} F_y(j,c)$ is the total number of records which involve China and another country as two actors. Note that here the connection strength is not normalized by the total occurrence of country $I$. Unlike the "two-way" spatial decay effect, here we concentrates on the "one-way" effect focusing on China as a target geographic entity, i.e., how important country $I$ is to China (based on the percentage of its co-occurrence with China) without considering how important China is for country $I$. In this way the time series study provides more valuable input from China's perspective.

To explore the changing dynamics of this pattern, we compute the yearly connection strength between China and the top 15 countries, represented as time series data. The following series provides an example series between US and China, which indicates that the connection strength is 0.162 in the year 1979 and 0.179 in 2013 (cf. Equation (1)):

- US [0.162, 0.174, 0.191, 0.193, 0.189, 0.189, 0.181, 0.177, 0.174, 0.169, 0.17, 0.165, 0.162, 0.157, 0.157, 0.161, 0.17, 0.165, 0.164, 0.165, 0.166, 0.16, 0.164, 0.16, 0.159, 0.155, 0.153, 0.151, 0.153, 0.156, 0.162, 0.169, 0.175, 0.178, 0.179]

- *Modeling and interpreting time series data*

As discussed in Section 1, ARIMA model can be applied for both stationary and non-stationary time series data. Due to its flexibility in data processing, this research constructed ARIMA models to better interpret the summarized time series. ARIMA model is generally referred to as an ARIMA($p,d,q$) model where three parameters $p$, $d$, and $q$ are non-negative integers. They refer to the autoregressive, integrated, and moving average parts of the model respectively, and are interpreted as follows:

- $p$: the autoregressive parameter indicates how much the output variable depends linearly on its own previous values (e.g., how much the value in 2010 depends on the years 2009, 2008, and etc.)
- $d$: the integrated parameter is the number of non-seasonal differences and long term trend. For instance, in the random walk model $Y(t) - Y(t-1) = \mu$ (where the average difference in Y is a constant, denoted by $\mu$), since it includes (only) a non-seasonal difference and a constant term, it is classified as an "ARIMA(0,1,0) model with constant.
- $q$: the order of lagged forecast errors in the prediction. For instance, if series $\mu_t$ can be represented by the weighted average of $q$ white noise patterns (Equation 1, where $\varepsilon_t$ are white noise series, $\theta_1 \ldots \theta_q$ are constants), then $\mu_t$ corresponds to ARIMA (0,0,$q$). q can be interpreted as a level of uncertainty in time series analysis:

$$\mu_t = \varepsilon_t + \theta_1 \varepsilon_{t-1} + \cdots \theta_q \varepsilon_{t-q} \quad (2)$$

The construction of ARIMA models provides quantitative evidences of how the inter-nation connection of China has changed upon time, and the fitted parameters can be applied for predictions and estimation of future patterns.

## 3. RESULTS AND DISCUSSIONS

ARIMA models introduced in Section 2 are constructed based on these time series. To test the effectiveness of the models, we utilized data from 1979-2012 as training set and the year 2013 as testing set for model validation. Table 1 presents the models and fitted results.

Table 1. ARIMA models and predicted results

| Country | FIPS Code[3] | ARIMA model | Fitted 2013 | Observed 2013 |
|---|---|---|---|---|
| United States | US | ARIMA(1,1,0) | 0.1798 | 0.1792 |
| Japan | JA | ARIMA(1,0,0) | 0.0924 | 0.0929 |
| Russia | RS | ARIMA(1,0,0) | 0.0819 | 0.0808 |
| South Korea | KS | ARIMA(0,1,1) | 0.0467 | 0.0465 |
| North Korea | KN | ARIMA(0,1,1) | 0.0421 | 0.0424 |
| United Kingdom | UK | ARIMA(1,0,2) | 0.0405 | 0.0414 |
| France | FR | ARIMA(0,1,0) | 0.029 | 0.029 |
| Iran | IR | ARIMA(0,1,0) | 0.0242 | 0.0238 |
| Pakistan | PK | ARIMA(2,0,0) | 0.0217 | 0.0236 |
| India | IN | ARIMA(1,1,0) | 0.0229 | 0.0227 |
| Australia | AS | ARIMA(1,1,0) | 0.0223 | 0.0219 |
| Vietnam | VM | ARIMA(1,2,0) | 0.0173 | 0.0193 |
| Germany | GM | ARIMA(0,1,1) | 0.0184 | 0.0184 |
| Philippines | RP | ARIMA(0,1,1) | 0.016 | 0.0152 |
| Canada | CA | ARIMA(2,1,0) | 0.0143 | 0.014 |

The fitted ARIMA models in Table 1 indicate interesting patterns. First, the non-zero *d* value (integrated parameter) for most countries indicates a fact that non-stationary long term trend exists in the connection between China and these countries. Figure 1 shows an example time series in South Korea showing a clear increasing trend (*d*=1).

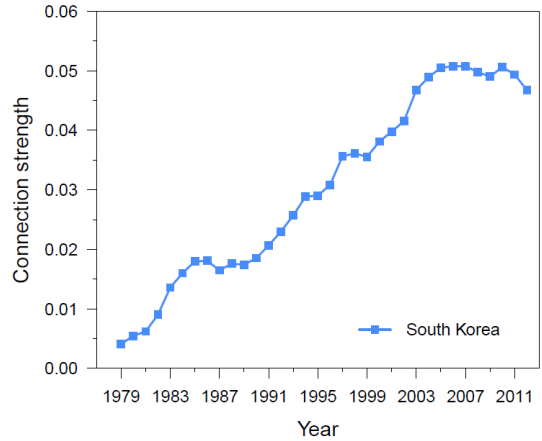

Figure 1. Yearly connection strength between China and South Korea

Moreover, Table 1 indicates that the 15 countries can be characterized into the following categories (Table 2):

Table 2. Categorized ARIMA models and countries

| | Characteristics | Countries |
|---|---|---|
| **Autoregressive models** $(p > 0, d = 0, q = 0)$ | The output variable depends linearly on its own previous values | JA, RS, PK |
| **Autoregressive integrated models** $(p > 0, d > 0, q = 0)$ | Autoregressive models with non-stationary behavior (e.g., long-term trend) | US, IN, AS, VM, CA |
| **Integrated moving average models** $(p = 0, d > 0, q > 0)$ | For moving average model, the output variable is conceptually a <u>linear regression</u> of the linear combination of q + 1 white noise variables. Integrated moving average models is MA model with non-stationary behavior | KS, KN, GM, RP |
| **Autoregressive Moving average models** $(p > 0, d = 0, q > $ | A combination of MA and AR models without non-stationary | UK |

---

[3] Federal Information Processing Standards (FIPS) are publicly announced standardizations developed by the United States federal government for use in computer systems.

| | | |
|---|---|---|
| **General integrated models (p = 0, d > 0, q = 0)** | The output variable depends only on the orders of non-stationary component | FR, IR |

Table 2 indicates varying patterns between different countries and China. For instance, the connection strength between China and Russia is fitted as a stationary process, in which the connection strength for a certain year auto-correlates with the value of the previous year; however, between China and France, the connection strength is a basic random walk model (ARIMA(0,1,0)) where the difference between two consecutive years can be modeled as a constant. To validate the models we also computed the predicted connection strength in 2013. The forecast accuracy level of the model is evaluated using mean absolute percentage error (MAPE):

$$MAPE = \frac{1}{n}\sum_{t=1}^{n} |\frac{Y_t - F_t}{Y_t}| \quad (2)$$

Where $n$ is the number of time points, $F_t$ is the forecast value at time t and $Y_t$ is the actual data. In Table 2 the average MAPE is 2.39%, which indicates a reliable model with low prediction error rate (<5%).

## 4. CONCLUSION

This paper employed the GDELT dataset to examine the connection between China and foreign countries based on time series analysis. We examined the effectiveness of ARIMA in predicting trends in long-term mass media data. Although ARIMA has been previously applied in political geography and communication fields, the application in determining inter-country relation in the big data era is limited. We also demonstrated the powerfulness of applying GDELT and big data techniques to investigate informative patterns for interdisciplinary researchers. This research does not aim to provide in-depth interpretation of the causes and consequences of these inter-nation events from a political perspective; instead, it proposed a method to discover the patterns that can provide insights in different research fields.

Potential future directions include extending this method to other countries to test its robustness. GDELT provides a rich data source to analyze inter-region relations at various spatial scales, such as investigating the connection between different provinces in China. Another valuable direction is to compare the performance of mass media and social media in characterizing urban-level patterns. Future study can also look into the correlation between connection strength and various demographic variables such as population, economic status and the tone of each event record.